\documentclass[english]{article}
\usepackage[english]{babel}
\usepackage[affil-it]{authblk}

\usepackage[numbers,sort]{natbib}
\usepackage[utf8]{inputenc} % allow utf-8 input
\usepackage[T1]{fontenc}    % use 8-bit T1 fonts

\usepackage{amsmath}
\usepackage{graphicx}
\usepackage{array}
\usepackage[colorlinks = true,
            linkcolor = blue,
            urlcolor  = blue,
            citecolor = blue,
            anchorcolor = blue]{hyperref}

\usepackage{amsfonts}       % blackboard math symbols
\usepackage[font=footnotesize,labelfont=bf]{caption}
\usepackage{hyphenat}
\usepackage{libertine}
\usepackage{fullpage}
\usepackage{siunitx}

\newcommand{\pointnet}{\texttt{PointNet}~}

\begin{document}

\title{Medical Application of Geometric Deep Learning for the Diagnosis of Glaucoma}

\author[1,$\star$]{Alexandre H. Thi\'{e}ry}
\author[2,3]{Fabian Braeu}
\author[4,5]{Tin A. Tun}
\author[3,4,5]{Tin Aung}
\author[2,5,6,$\star$]{Micha\"el J.A. Girard}

\affil[1]{Department of Statistics and Data Science, National University of Singapore, Singapore}
\affil[2]{Ophthalmic Engineering and Innovation Laboratory, Singapore Eye Research Institute, Singapore National Eye Centre, Singapore}
\affil[3]{Yong Loo Lin School of Medicine, National University of Singapore, Singapore}
\affil[4]{Singapore Eye Research Institute, Singapore National Eye Centre, Singapore}
\affil[5]{Duke-NUS Graduate Medical School, Singapore}
\affil[6]{Institute for Molecular and Clinical Ophthalmology, Basel, Switzerland}
\bigskip
\affil[$\star$]{Corresponding authors}

\maketitle

\begin{abstract}
\noindent
{\bf Purpose:} {\bf (1)} To assess the performance of geometric deep learning (\texttt{PointNet}) in diagnosing glaucoma from a single optical coherence tomography (OCT) 3D scan of the optic nerve head (ONH); {\bf (2)} To compare its performance to that obtained with a standard 3D convolutional neural network (CNN), and with a gold-standard glaucoma parameter, i.e. retinal nerve fiber layer (RNFL) thickness. \\

\noindent
{\bf Methods:} 3D raster scans of the ONH were acquired with Spectralis OCT (Heidelberg Engineering, Heidelberg, Germany) for $477$ glaucoma and $2,296$ non-glaucoma subjects at the Singapore National Eye Centre. All volumes were automatically segmented using deep learning to identify $7$ major neural and connective tissues including the RNFL, the prelamina, and the lamina cribrosa (LC). Each ONH was then represented as a 3D point cloud with $1,000$ points chosen randomly from all tissue boundaries. To simplify the problem, all ONH point clouds were aligned with respect to the plane and center of Bruch’s membrane opening. Geometric deep learning (\texttt{PointNet}) was then used to provide a glaucoma diagnosis from a single OCT point cloud. The performance of our approach (reported using the area under the receiver operating characteristic curve or AUC) was compared to that obtained with a 3D CNN, and with RNFL thickness.\\

\noindent
{\bf Results:} \pointnet was able to provide a robust glaucoma diagnosis solely from the ONH  represented as a 3D point cloud (AUC=$0.95 \pm 0.01$). The performance of PointNet was superior to that obtained with a standard 3D CNN (AUC=$0.87 \pm 0.02$) and with that obtained from RNFL thickness alone (AUC=$0.80 \pm 0.03$).\\

\noindent
{\bf Discussion:} We provide a proof-of-principle for the application of geometric deep learning in the field of glaucoma. Our technique requires significantly less information as input to perform better than a 3D CNN, and with an AUC superior to that obtained from RNFL thickness alone. Geometric deep learning may have wide applicability in the field of Ophthalmology.\\

\noindent
{\bf Keywords:} PointNet, Glaucoma, Optical Coherence Tomography, Optic Nerve Head, Artificial Intelligence
\end{abstract}

\section{Introduction}
With the development and progression of glaucoma, the optic nerve head (ONH) typically exhibits complex neural- and connective-tissue structural changes including, but not limited to: thinning of the retinal nerve fiber layer (RNFL) and the ganglion cell complex (GCC) layer \cite{kanamori2003evaluation, mwanza2012glaucoma}, changes in lamina cribrosa (LC) shape/depth/curvature \cite{kim2018lamina,downs2017lamina}, and posterior bowing of the peripapillary sclera \cite{wang2022peripapillary,wang2020peripapillary}. Clinically, optical coherence tomography (OCT) is the mainstay of imaging to observe such changes \cite{geevarghese2021optical}, however, signal interpretation (by humans or machines) for glaucoma diagnosis/prognosis remains a challenge. 
Recently, a growing number of artificial intelligence (AI) studies have proposed to use deep learning algorithms to provide a robust glaucoma diagnosis from a single OCT scan of the ONH or of the macula. Such applications could have excellent clinical value by e.g. reducing the number of tests needed to confirm a glaucoma diagnosis. Some of these algorithms were directly applied to the raw OCT scans \cite{maetschke2019feature,ran2019detection,russakoff20203d}, while others, first simplified the images to only a few classes (or colors) by highlighting relevant tissue structures \cite{panda2022describing}. Most algorithms achieved good to excellent performance. However, such algorithms need to be able to handle a considerable amount of information (e.g. voxel intensities distributed on 3D grids) that could be heavily corrupted by noise, image artifacts, and 3D image orientation issues, thus limiting their ease-of-use and deployability.\\

To this end, a family of algorithms, known as \pointnet \cite{qi2017pointnet}, fitting under the category of geometric deep learning, has been proposed to solve classification problems from structures represented as 3D point clouds (such as those in medical imaging \cite{gutierrez2018deep}), with excellent performance. Because such point clouds do not need to be dense and falling onto regular grids, this intrinsically means the amount of information needed to make e.g. a diagnosis could be significantly reduced, thus reducing the ‘black-box’ element of AI. For our glaucoma diagnosis application, the ONH can simply be thought of as a complex 3D structure that can be represented by a cloud of points as has been routinely performed in 3D histomorphometric and finite element studies \cite{yang20183d,sigal2004finite,jin2020effect}.\\

In this study, we aimed to apply a geometric deep learning solution (\texttt{PointNet}) to provide a robust glaucoma diagnosis from a single OCT scan of the ONH. Each OCT scan was first pre-processed and each ONH was represented as a 3D point cloud, thus limiting the amount of information to be processed by several orders of magnitude. Our approach was compared with a standard 3D convolutional neural network (CNN), and its performance compared with that from RNFL thickness alone (current gold standard). It may have the potential to considerably simplify diagnosis and prognosis applications in the field of glaucoma.

\section{Methods}

\subsection{Patient Recruitment}
A total of $2,773$ subjects ($477$ glaucoma and $2,296$ non-glaucoma) were recruited at the Singapore National Eye Center (SNEC, Singapore; see Table \ref{tab:summary} for further demographic details). All subjects gave written informed consent. The study adhered to the tenets of the Declaration of Helsinki and was approved by the institutional review board of the respective hospital. Subjects with intraocular pressure (IOP) less than $21$ mmHg, healthy optic discs with a vertical cup-disc ratio (VCDR) less than or equal to $0.5$, and normal visual fields tests were considered as non-glaucoma, whereas subjects with glaucomatous optic neuropathy, VCDR $\geq 0.7$, and/or neuroretinal rim narrowing with repeatable glaucomatous visual field defects were considered as glaucoma. Subjects with corneal abnormalities that have the potential to preclude the quality of the scans were excluded from the study.

\begin{table}[h]
    \centering
    \begin{tabular}{|m{4em}|m{4em}|m{3em}|m{6.5em}|m{4em}|m{6.5em}|m{4em}|}
    \hline
    %\multicolumn{7}{|c|}{Country List}\\
    Institute & Age & Sex (\%Male) & Non-Glaucoma Patients & Glaucoma Patients & Non-Glaucoma Scans & Glaucoma Scans\\
    \hline
    SNEC & $60.5\pm9.0$ & $51\%$ & $2,296$ & $477$ & $3,897$ & $873$\\
    \hline
    \multicolumn{3}{|c|}{Total} & \multicolumn{2}{|c|}{$2,773$} & \multicolumn{2}{|c|}{$4,770$} \\
    \hline
    \end{tabular}
    \caption{Summary of patient information.}
    \label{tab:summary}
\end{table}

\subsection{Optical Coherence Tomography Imaging}
Spectral-domain OCT imaging (Spectralis; Heidelberg Engineering, Heidelberg, Germany) was performed on seated subjects under dark room conditions after dilation with tropicamide $1\%$ solution. Images were acquired from either both eyes or one eye of each subject. Each OCT volume consisted of 97 serial horizontal Bscans ($\sim \SI{30}{\micro\metre}$ distance between B-scans; $384$ A-scans per B-scan; $20 \times$ signal averaging; axial resolution: $\sim \SI{3.87}{\micro\metre}$) that covered a rectangular area of $15^{\circ} \times 10^{\circ}$ centered on the ONH. The eye tracking and enhanced depth imaging modalities of the Spectralis were used during image acquisition. In total, we obtained $4,770$ scans ($873$ glaucoma and $3,897$ non-glaucoma scans).   

\subsection{Automated Segmentation of OCT Images}
\begin{figure}[ht]
\centering
\includegraphics[width=0.8 \textwidth]{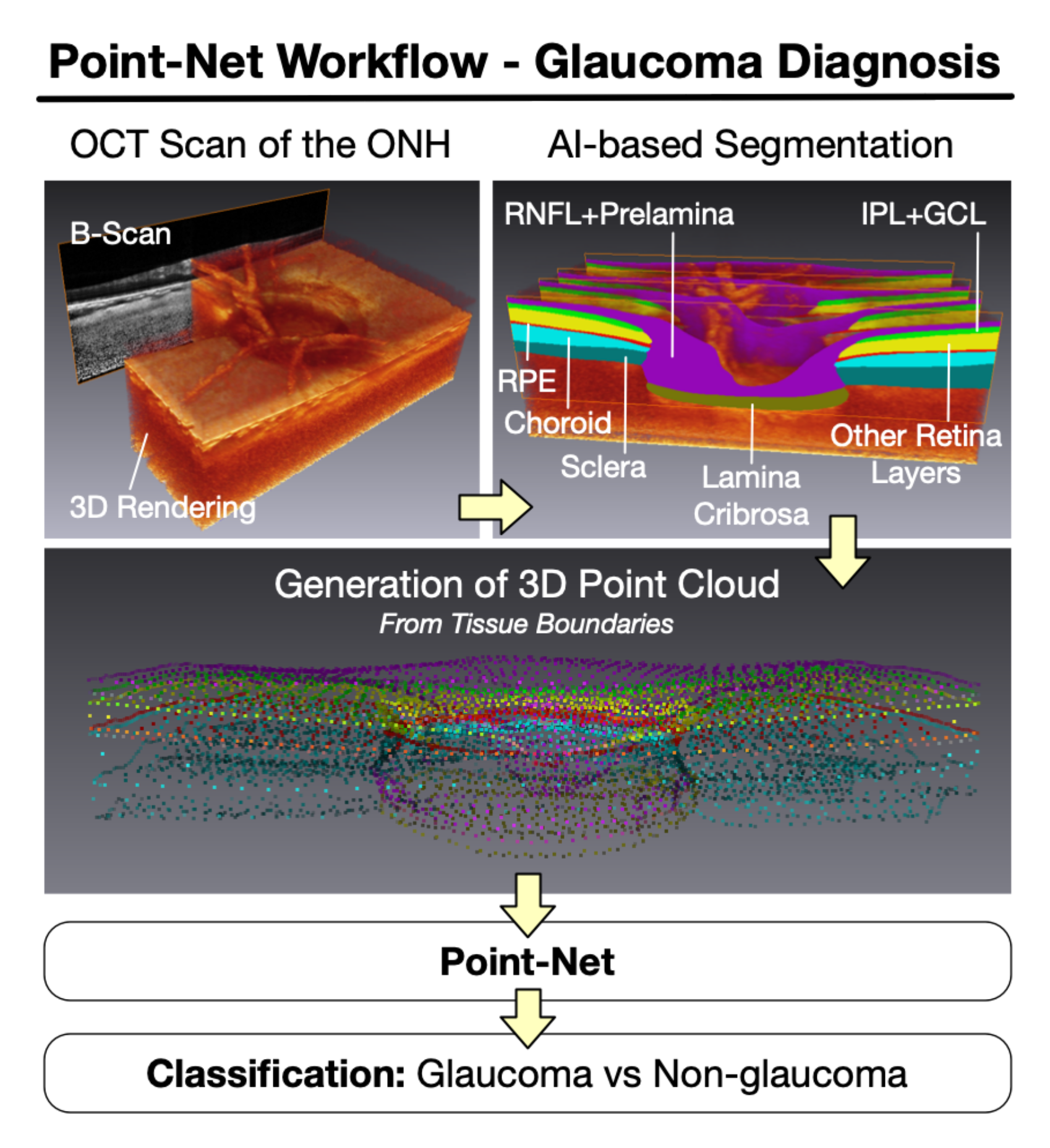}
\caption{\pointnet Workflow: each OCT scan of the ONH is first segmented using deep learning to identify the following tissue structures: RNFL+prelamina, IPL+GCL, all other retina layers, RPE, choroid, peripapillary sclera, and lamina cribrosa. A 3D point cloud is then generated strictly from the tissue boundaries. The 3D point cloud is ultimately passed through our \pointnet network to produce a glaucoma diagnosis.}
\label{fig:workflow}
\end{figure}

Since \pointnet requires each ONH to be described as a point cloud, it was first necessary for us to identify (or highlight) the major neural and connective tissues that are involved in glaucoma pathogenesis. To this end, we used the software \texttt{REFLECTIVITY}~ (Abyss Processing Pte Ltd, Singapore) to automatically segment the following tissue groups: (1) the retinal nerve fiber layer (RNFL) and the prelamina; (2) the ganglion cell layer and the inner plexiform layer (GCL+IPL); (3) all other retinal layers; (4) the retinal pigment epithelium (RPE) with Bruch’s membrane (BM); (5) the choroid; (6) the peripapillary sclera including the scleral flange; and (7) the lamina cribrosa (LC). All segmented tissues can be observed in Figure \ref{fig:workflow}. Note that \texttt{REFLECTIVITY}~ was developed from advances in AI-based ONH segmentation as described in our previous publications \cite{devalla2018drunet,devalla2020towards}. It is also important to point out that \texttt{REFLECTIVITY}~ cannot identify the "true" posterior boundaries of the peripapillary sclera and of the LC, but instead provides the OCT-visible portions of those 2 tissues. The AI-based segmentation process assigned a label to each voxel of each 3D OCT scan to indicate the tissue class.

\subsection{Point cloud generation}
Once each voxel of each 3D OCT scan was assigned a label, the voxels situated at the boundaries between two different tissue groups were automatically identified. The following N=8 boundaries of interest were identified: anterior \& posterior boundaries of the RNFL+prelamina, and the posterior boundaries of the IPL+GLC, other retina layers, RPE, choroid, sclera, and LC. On average, a total of $\sim 50,000$ boundary voxels of interest were extracted from each 3D OCT scan. The 3D coordinates as well as the boundary class (expressed as a label varying between $1$ and $N=8$) of each boundary voxel of interest was recorded for further processing. The 3D coordinates were expressed within an [$x,y,z$] cartesian coordinate system with origin situated at the center of the Bruch’s membrane opening (BMO) circle, and such that the BMO circle lies in the horizontal [$x=0$, $y=0$] plane.

\subsection{PointNet for Glaucoma Diagnosis}

In this work, we aimed to design a binary classifier to identify whether a given ONH would be classified as glaucoma or non-glaucoma. For designing such a classifier, we opted to use a simple and robust solution that produces a probabilistic forecast based only on the geometric properties of the ONH.\\

Geometric Deep Learning \cite{bronstein2017geometric} is an emerging field of AI that proposes inductive biases and network architectures that can efficiently process data structures such as grids, graphs and cloud of points while respecting their intrinsic symmetries and invariances \cite{wu2020comprehensive}. In this study, we have leveraged the recently proposed \pointnet neural architecture \cite{qi2017pointnet}. This deep neural network has been especially designed to process point-clouds, i.e. an unordered set of points. A \pointnet takes a point-cloud as input and provides a unified architecture for applications ranging from object classification, part segmentation, to scene semantic parsing. Although simple, it has been empirically demonstrated that the \pointnet architecture  exhibits a performance on par, or even better, than the state-of-the-art \cite{qi2017pointnet}.\\

For classifying a 3D OCT scan as glaucoma or non-glaucoma, we used a \pointnet that processed a point-cloud of size $S=1,000$. For this purpose, out of the typically much larger set of points of interest extracted from each OCT scan, a subset of S=1,000 points were randomly selected. There are two main reasons for this approach: (1) we have empirically observed that this random sub-sampling procedure improved the robustness of the method and helped mitigate overfitting issues; and (2) among all the points of interests extracted from each 3D OCT scan, there is typically a large amount of redundancy. Reducing the point-clouds to a random subset of $S=1,000$ points helped improve the computational efficiency of the method and led to more robust predictions. The point location (i.e. 3 dimensional coordinates) as well as its boundary class (represented as a one-hot-encoded vector of dimension $N=8$) were concatenated into a vector of dimension $D=8+3=11$. In summary, the \pointnet was designed to process point-clouds of size $S=1,000$ and dimension $D=11$.\\

The dataset of OCT scans was split into training ($70\%$), validation ($15\%$), and test ($15\%$) sets, respectively. The split was performed in such a way that scans from the same subject did not exist in different sets. Furthermore, the proportion of glaucoma scans was identical in all the splits. Our network was then trained on a Nvidia 1080Ti GPU card until optimum performance was reached in the validation set in about $100$ epochs. For this purpose, the standard cross-entropy loss was minimized with the \texttt{ADAM}~ optimizer \cite{kingma2014adam}. During the training process, each time an OCT scan was processed (i.e. once during each epoch), a newly generated subset of S=1,000 points was selected and fed into the \pointnet architecture. For enhanced robustness and to further enrich the training set, subsequent to the random subsampling process we also used a data-augmentation scheme that applied random rigid transformations to the subsampled point-clouds. When applying a rigid transformation to such a point-cloud, the rigid transformation was only applied to the spatial coordinates (i.e. the first 3 coordinates). To evaluate the performance of our method, we reported the area under the receiver operating characteristic curve (ROC-AUC) with uncertainty estimates obtained from a $5$-fold cross-validation process.\\

\begin{figure}[ht]
\centering
\includegraphics[width=0.6 \textwidth]{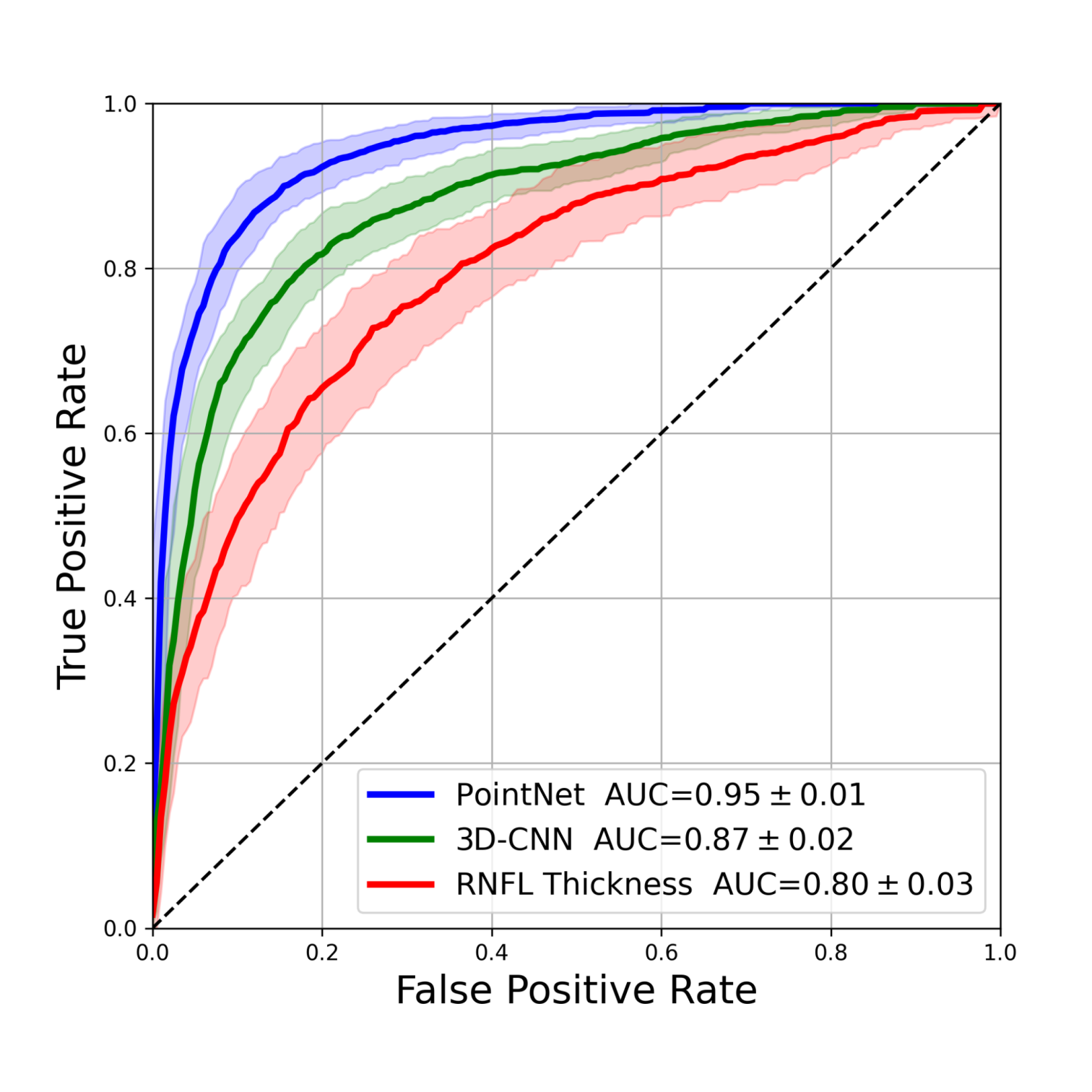}
\caption{The AUCs were found to be $0.80\pm0.03$ for RNFL thickness, $0.87\pm0.02$ for the 3D CNN approach, and $0.95\pm0.01$ for the \pointnet method.}
\label{fig:roc_auc}
\end{figure}

\subsection{Comparisons to 3D Convolutional Neural Network}
To compare the performance of our \pointnet with a Gold-standard approach we trained a 3D convolutional neural network (CNN) \cite{maetschke2019feature}. For this purpose, each volume was linearly resampled to isotropic dimensions $128 \times 128 \times 128$ voxels. The resized OCT scans were fed into a CNN composed of five 3D-convolutional layers, alternated with batch-normalization operations and ReLU non-linearities. After a final global average pooling layer, a softmax operation was used to produce a probabilistic binary classification estimate. The standard cross-entropy loss was minimized with the \texttt{ADAM} optimizer. As employed during the training of the \pointnet, the dataset of OCT scans was split into training ($70\%$), validation ($15\%$), and test ($15\%$) sets, respectively. It was important to use data-augmentation during the training process: random translations and rotations ($\pm 15$ degrees), left-right flips, additions of Gaussian noise to the voxel intensities. We used early stopping and selected the network with the highest validation ROC-AUC during training. We reported the classification ROC-AUC with uncertainty estimates obtained from a $5$-fold cross-validation process. 

\subsection{Comparisons to RNFL Thickness}
To compare the performance of \pointnet to that from a gold standard glaucoma parameter \cite{kanamori2003evaluation}, we extracted, from the 3D segmentation of the OCT scans, the average RNFL thickness at a distance of $1.4$ times the BMO radius. RNFL thickness was computed from our automated segmentations as the minimum distance between the anterior and posterior boundaries of the RNFL. We then reported the diagnostic power as quantified by the AUC. Since the RNFL thickness is a scalar parameter, no classification algorithm was needed to compute the AUC.

\subsection{Results}
We evaluated the classification performance of {\bf (1)} RNFL thickness as a gold standard glaucoma parameter; of {\bf (2)} the 3D CNN approach; and of {\bf (3)} our proposed \pointnet method by performing a $5$-fold cross-validation study. The three methods were evaluated on the same splits of the data. The AUCs were found to be $0.80\pm0.03$ for RNFL Thickness, $0.87\pm0.02$ for the 3D CNN approach, and $0.95\pm0.01$ for the \pointnet method (Figure \ref{fig:roc_auc}).

\section{Discussion}
In this study, we proposed a relatively simple AI approach based on geometric deep learning to provide a robust glaucoma diagnosis from a single OCT scan of the ONH. Each ONH was first pre-processed as a 3D point cloud in order to represent all major neural- and connective-tissue boundaries. Overall, our proposed approach required significantly less information to perform glaucoma classification than other 3D AI approaches. It also performed better than a standard 3D CNN (applied to raw OCT scans), and by taking into account information about both ONH neural and connective tissue layers, it yielded an AUC higher than that obtained from RNFL thickness alone.\\

In this study, we found that geometric deep learning (\texttt{PointNet}) was well adapted to perform glaucoma classification with ONH tissues represented as point clouds. In this study, a given ONH was represented with $1,000$ data points (as a point cloud), which is significantly less than the $\sim18,000,000$ data points (or voxels) needed to represent an 3D OCT scan. In other words, we reduced the size of our input by $4$ orders or magnitude. Dealing with smaller inputs may ultimately allow us to reduce the "black-box" effect of AI, and allow us to better identify the complex 3D structural and biomechanical signature of the glaucomatous ONH \cite{sigal2009biomechanics}.\\

It is interesting to note that our \pointnet performed better than a standard 3D CNN on the exact same dataset (AUC=$0.95\pm0.01$ for \pointnet vs AUC=$0.87\pm0.02$ for 3D CNN). This may not be surprising as 3D OCT scans typically exhibit a considerable amount of noise and artifacts that would have been otherwise eliminated through a pre-processing segmentation step, as proposed herein. In addition, our \pointnet was able to focus on the most important ONH structural features (such as tissue boundaries, and implicitly, tissue thicknesses), which would be considerably more challenging to fully exploit for a 3D CNN.\\

In this study, we found that \pointnet was able to provide a higher diagnostic accuracy (AUC=$0.95\pm0.01$) as compared to that obtained from a gold standard glaucoma parameter, i.e. RNFL thickness (AUC=$0.80\pm0.03$). This result may not be surprising as the ONH do not only exhibit neural tissue changes, but also connective tissue changes, such as bending of the peripapillary sclera \cite{wang2022peripapillary}, and changes in LC morphology and pore shape/pathway \cite{wang2018tortuous,shoji2017glaucomatous}. \pointnet has the advantage of capturing some of these features while minimizing the total amount of information required to establish a diagnosis. \pointnet may also help us identify the contribution of each individual neural or connective tissue for the diagnosis or prognosis of glaucoma.\\

Geometric deep learning may have wide applicability in the field of Ophthalmology. It is relatively attractive for its ease of use, ease (and speed) of training, and considerably smaller and simpler input size. While our first application targeted glaucoma, geometric deep learning could also be used for the diagnosis and prognosis of other optic neuropathies \cite{girard20213d}, and for a wide range of corneal \cite{dos2019corneanet} and retinal disorders \cite{schmidt2021ai}.\\

In this study, several limitations warrant further discussion. First, our approach was only tested with one OCT device (Spectralis OCT), and for one Singapore population. Second, we did not consider all structural landmarks that could have improved the diagnosis of glaucoma such as e.g. the 3D configuration of the central retinal vessels \cite{panda2022three}, or the presence of peripapillary atrophy \cite{wang2021optic}; neither did we consider other optical properties \cite{leung2022diagnostic}. Third, our non-glaucoma population did not include other major optic neuropathies. The inclusion of such cases could be critical for clinical translation \cite{al2022artificial}. Fourth, the segmentation of the posterior LC and peripapillary sclera boundaries was artificial and was solely based on the amount of visible signal in the OCT scans, which most likely does not coincide with the "true" anatomical landmarks \cite{girard2011shadow}. This may have influenced our diagnostic powers.\\

In conclusion, we provide herein a proof-of-principle for the application of geometric deep learning in the field of Ophthalmology with a special emphasis on glaucoma diagnosis. We found that our technique required significantly less data as input to perform better than a 3D CNN, and with an AUC superior to that obtained from RNFL thickness alone. Geometric deep learning may have wide applicability in the field of Ophthalmology, and it should be explored for other pathologies.

\section*{Acknowledgments}
We acknowledge funding from {\bf (1)} the donors of the National Glaucoma Research, a program of the BrightFocus Foundation, for support of this research (G2021010S [MJAG]); {\bf (2)} SingHealth Duke-NUS Academic Medicine Research Grant (SRDUKAMR21A6 [MJAG]); {\bf (3)}  Singapore MOE Tier 1 grant (R155-000-228-114) [AHT]; {\bf (4)} the "Retinal Analytics through Machine learning aiding Physics (RAMP)" project supported by the National Research Foundation, Prime Minister's Office, Singapore under its Intra-Create Thematic Grant "Intersection Of Engineering And Health" - NRF2019-THE002-0006 [MJAG/AT].

\section*{Disclosure}
AHT and MJAG are the co-founders of the AI start-up company Abyss Processing Pte Ltd that provides 3D AI solutions for glaucoma diagnosis and prognosis  	

\bibliographystyle{unsrt}
\bibliography{sample}

\begin{thebibliography}{10}

\bibitem{kanamori2003evaluation}
Akiyasu Kanamori, Makoto Nakamura, Michael~FT Escano, Ryu Seya, Hidetaka Maeda,
  and Akira Negi.
\newblock Evaluation of the glaucomatous damage on retinal nerve fiber layer
  thickness measured by optical coherence tomography.
\newblock {\em American journal of ophthalmology}, 135(4):513--520, 2003.

\bibitem{mwanza2012glaucoma}
Jean-Claude Mwanza, Mary~K Durbin, Donald~L Budenz, Fouad~E Sayyad, Robert~T
  Chang, Arvind Neelakantan, David~G Godfrey, Randy Carter, and Alan~S
  Crandall.
\newblock Glaucoma diagnostic accuracy of ganglion cell--inner plexiform layer
  thickness: comparison with nerve fiber layer and optic nerve head.
\newblock {\em Ophthalmology}, 119(6):1151--1158, 2012.

\bibitem{kim2018lamina}
Jeong-Ah Kim, Tae-Woo Kim, Robert~N Weinreb, Eun~Ji Lee, Micha{\"e}l~JA Girard,
  and Jean~Martial Mari.
\newblock Lamina cribrosa morphology predicts progressive retinal nerve fiber
  layer loss in eyes with suspected glaucoma.
\newblock {\em Scientific reports}, 8(1):1--10, 2018.

\bibitem{downs2017lamina}
J~Crawford Downs and Christopher~A Girkin.
\newblock Lamina cribrosa in glaucoma.
\newblock {\em Current opinion in ophthalmology}, 28(2):113, 2017.

\bibitem{wang2022peripapillary}
Xiaofei Wang, Tin~A Tun, Monisha~Esther Nongpiur, Hla~M Htoon, Yih~Chung Tham,
  Nicholas~G Strouthidis, Tin Aung, Ching-Yu Cheng, and Michael~JA Girard.
\newblock Peripapillary sclera exhibits a v-shaped configuration that is more
  pronounced in glaucoma eyes.
\newblock {\em British Journal of Ophthalmology}, 106(4):491--496, 2022.

\bibitem{wang2020peripapillary}
Ya~Xing Wang, Hongli Yang, Haomin Luo, Seung~Woo Hong, Stuart~K Gardiner,
  Jin~Wook Jeoung, Christy Hardin, Glen~P Sharpe, Kouros Nouri-Mahdavi, Joseph
  Caprioli, et~al.
\newblock Peripapillary scleral bowing increases with age and is inversely
  associated with peripapillary choroidal thickness in healthy eyes.
\newblock {\em American journal of ophthalmology}, 217:91--103, 2020.

\bibitem{geevarghese2021optical}
Alexi Geevarghese, Gadi Wollstein, Hiroshi Ishikawa, and Joel~S Schuman.
\newblock Optical coherence tomography and glaucoma.
\newblock {\em Annual review of vision science}, 7:693--726, 2021.

\bibitem{maetschke2019feature}
Stefan Maetschke, Bhavna Antony, Hiroshi Ishikawa, Gadi Wollstein, Joel
  Schuman, and Rahil Garnavi.
\newblock A feature agnostic approach for glaucoma detection in oct volumes.
\newblock {\em PloS one}, 14(7):e0219126, 2019.

\bibitem{ran2019detection}
An~Ran Ran, Carol~Y Cheung, Xi~Wang, Hao Chen, Lu-yang Luo, Poemen~P Chan,
  Mandy~OM Wong, Robert~T Chang, Suria~S Mannil, Alvin~L Young, et~al.
\newblock Detection of glaucomatous optic neuropathy with spectral-domain
  optical coherence tomography: a retrospective training and validation
  deep-learning analysis.
\newblock {\em The Lancet Digital Health}, 1(4):e172--e182, 2019.

\bibitem{russakoff20203d}
Daniel~B Russakoff, Suria~S Mannil, Jonathan~D Oakley, An~Ran Ran, Carol~Y
  Cheung, Srilakshmi Dasari, Mohammed Riyazzuddin, Sriharsha Nagaraj, Harsha~L
  Rao, Dolly Chang, et~al.
\newblock A 3d deep learning system for detecting referable glaucoma using full
  oct macular cube scans.
\newblock {\em Translational Vision Science \& Technology}, 9(2):12--12, 2020.

\bibitem{panda2022describing}
Satish~K Panda, Haris Cheong, Tin~A Tun, Sripad~K Devella, Vijayalakshmi
  Senthil, Ramaswami Krishnadas, Martin~L Buist, Shamira Perera, Ching-Yu
  Cheng, Tin Aung, et~al.
\newblock Describing the structural phenotype of the glaucomatous optic nerve
  head using artificial intelligence.
\newblock {\em American journal of ophthalmology}, 236:172--182, 2022.

\bibitem{qi2017pointnet}
Charles~R Qi, Hao Su, Kaichun Mo, and Leonidas~J Guibas.
\newblock Pointnet: Deep learning on point sets for 3d classification and
  segmentation.
\newblock In {\em Proceedings of the IEEE conference on computer vision and
  pattern recognition}, pages 652--660, 2017.

\bibitem{gutierrez2018deep}
Benjam{\'\i}n Guti{\'e}rrez-Becker and Christian Wachinger.
\newblock Deep multi-structural shape analysis: application to neuroanatomy.
\newblock In {\em International Conference on Medical Image Computing and
  Computer-Assisted Intervention}, pages 523--531. Springer, 2018.

\bibitem{yang20183d}
Hongli Yang, Juan Reynaud, Howard Lockwood, Galen Williams, Christy Hardin,
  Luke Reyes, Stuart~K Gardiner, and Claude~F Burgoyne.
\newblock 3d histomorphometric reconstruction and quantification of the optic
  nerve head connective tissues.
\newblock In {\em Glaucoma}, pages 207--267. Springer, 2018.

\bibitem{sigal2004finite}
Ian~A Sigal, John~G Flanagan, Inka Tertinegg, and C~Ross Ethier.
\newblock Finite element modeling of optic nerve head biomechanics.
\newblock {\em Investigative ophthalmology \& visual science},
  45(12):4378--4387, 2004.

\bibitem{jin2020effect}
Yuejiao Jin, Xiaofei Wang, Sylvi Febriana~Rachmawati Irnadiastputri,
  Rosmin~Elsa Mohan, Tin Aung, Shamira~A Perera, Craig Boote, Jost~B Jonas,
  Leopold Schmetterer, and Micha{\"e}l J~A Girard.
\newblock Effect of changing heart rate on the ocular pulse and dynamic
  biomechanical behavior of the optic nerve head.
\newblock {\em Investigative ophthalmology \& visual science}, 61(4):27--27,
  2020.

\bibitem{devalla2018drunet}
Sripad~Krishna Devalla, Prajwal~K Renukanand, Bharathwaj~K Sreedhar, Giridhar
  Subramanian, Liang Zhang, Shamira Perera, Jean-Martial Mari, Khai~Sing Chin,
  Tin~A Tun, Nicholas~G Strouthidis, et~al.
\newblock Drunet: a dilated-residual u-net deep learning network to segment
  optic nerve head tissues in optical coherence tomography images.
\newblock {\em Biomedical optics express}, 9(7):3244--3265, 2018.

\bibitem{devalla2020towards}
Sripad~Krishna Devalla, Tan~Hung Pham, Satish~Kumar Panda, Liang Zhang,
  Giridhar Subramanian, Anirudh Swaminathan, Chin~Zhi Yun, Mohan Rajan, Sujatha
  Mohan, Ramaswami Krishnadas, et~al.
\newblock Towards label-free 3d segmentation of optical coherence tomography
  images of the optic nerve head using deep learning.
\newblock {\em Biomedical optics express}, 11(11):6356--6378, 2020.

\bibitem{bronstein2017geometric}
Michael~M Bronstein, Joan Bruna, Yann LeCun, Arthur Szlam, and Pierre
  Vandergheynst.
\newblock Geometric deep learning: going beyond euclidean data.
\newblock {\em IEEE Signal Processing Magazine}, 34(4):18--42, 2017.

\bibitem{wu2020comprehensive}
Zonghan Wu, Shirui Pan, Fengwen Chen, Guodong Long, Chengqi Zhang, and S~Yu
  Philip.
\newblock A comprehensive survey on graph neural networks.
\newblock {\em IEEE transactions on neural networks and learning systems},
  32(1):4--24, 2020.

\bibitem{kingma2014adam}
Diederik~P Kingma and Jimmy Ba.
\newblock Adam: A method for stochastic optimization.
\newblock {\em arXiv preprint arXiv:1412.6980}, 2014.

\bibitem{sigal2009biomechanics}
Ian~A Sigal and C~Ross Ethier.
\newblock Biomechanics of the optic nerve head.
\newblock {\em Experimental eye research}, 88(4):799--807, 2009.

\bibitem{wang2018tortuous}
Bo~Wang, Katie~A Lucy, Joel~S Schuman, Ian~A Sigal, Richard~A Bilonick, Chen
  Lu, Jonathan Liu, Ireneusz Grulkowski, Zachary Nadler, Hiroshi Ishikawa,
  et~al.
\newblock Tortuous pore path through the glaucomatous lamina cribrosa.
\newblock {\em Scientific reports}, 8(1):1--7, 2018.

\bibitem{shoji2017glaucomatous}
Takuhei Shoji, Hiroto Kuroda, Masayuki Suzuki, Hisashi Ibuki, Makoto Araie, and
  Shin Yoneya.
\newblock Glaucomatous changes in lamina pores shape within the lamina cribrosa
  using wide bandwidth, femtosecond mode-locked laser oct.
\newblock {\em Plos one}, 12(7):e0181675, 2017.

\bibitem{girard20213d}
Micha{\"e}l~JA Girard, Satish~K Panda, Tin~Aung Tun, Elisabeth~A Wibroe,
  Raymond~P Najjar, Aung Tin, Alexandre~H Thi{\'e}ry, Steffen Hamann, Clare
  Fraser, and Dan Milea.
\newblock 3d structural analysis of the optic nerve head to robustly
  discriminate between papilledema and optic disc drusen.
\newblock {\em arXiv preprint arXiv:2112.09970}, 2021.

\bibitem{dos2019corneanet}
Valentin~Aranha Dos~Santos, Leopold Schmetterer, Hannes Stegmann, Martin
  Pfister, Alina Messner, Gerald Schmidinger, Gerhard Garhofer, and Ren{\'e}~M
  Werkmeister.
\newblock Corneanet: fast segmentation of cornea oct scans of healthy and
  keratoconic eyes using deep learning.
\newblock {\em Biomedical optics express}, 10(2):622--641, 2019.

\bibitem{schmidt2021ai}
Ursula Schmidt-Erfurth, Gregor~S Reiter, Sophie Riedl, Philipp Seeb{\"o}ck,
  Wolf-Dieter Vogl, Barbara~A Blodi, Amitha Domalpally, Amani Fawzi, Yali Jia,
  David Sarraf, et~al.
\newblock Ai-based monitoring of retinal fluid in disease activity and under
  therapy.
\newblock {\em Progress in retinal and eye research}, page 100972, 2021.

\bibitem{panda2022three}
Satish~K Panda, Haris Cheong, Tin~A Tun, Thanadet Chuangsuwanich, Aiste
  Kadziauskiene, Vijayalakshmi Senthil, Ramaswami Krishnadas, Martin~L Buist,
  Shamira Perera, Ching-Yu Cheng, et~al.
\newblock The three-dimensional structural configuration of the central retinal
  vessel trunk and branches as a glaucoma biomarker.
\newblock {\em American Journal of Ophthalmology}, 2022.

\bibitem{wang2021optic}
Ya~Xing Wang, Songhomitra Panda-Jonas, and Jost~B Jonas.
\newblock Optic nerve head anatomy in myopia and glaucoma, including
  parapapillary zones alpha, beta, gamma and delta: histology and clinical
  features.
\newblock {\em Progress in retinal and eye research}, 83:100933, 2021.

\bibitem{leung2022diagnostic}
Christopher Kai~Shun Leung, Alexander Ka~Ngai Lam, Robert~Neal Weinreb, David~F
  Garway-Heath, Marco Yu, Philip~Yawen Guo, Vivian Sheung~Man Chiu, Kelvin
  Ho~Nam Wan, Mandy Wong, Ken~Zhongheng Wu, et~al.
\newblock Diagnostic assessment of glaucoma and non-glaucomatous optic
  neuropathies via optical texture analysis of the retinal nerve fibre layer.
\newblock {\em Nature Biomedical Engineering}, pages 1--12, 2022.

\bibitem{al2022artificial}
Lama~A Al-Aswad, Rithambara Ramachandran, Joel~S Schuman, Felipe Medeiros,
  Malvina~B Eydelman, et~al.
\newblock Artificial intelligence for glaucoma: Creating and implementing ai
  for disease detection and progression.
\newblock {\em Ophthalmology Glaucoma}, 2022.

\bibitem{girard2011shadow}
Micha{\"e}l~JA Girard, Nicholas~G Strouthidis, C~Ross Ethier, and Jean~Martial
  Mari.
\newblock Shadow removal and contrast enhancement in optical coherence
  tomography images of the human optic nerve head.
\newblock {\em Investigative ophthalmology \& visual science},
  52(10):7738--7748, 2011.

\end{thebibliography}

\end{document}